\documentclass{article}
\usepackage{ijcai22}

\usepackage{times}
\usepackage{soul}
\usepackage{url}
\usepackage[hidelinks]{hyperref}
\usepackage[utf8]{inputenc}
\usepackage[small]{caption}
\usepackage{graphicx}
\usepackage{amsmath}
\usepackage{amsthm}
\usepackage{booktabs}
\usepackage{algorithm}
\usepackage{algorithmic}
\usepackage{diagbox}
\title{Enhancing Clean Label Backdoor Attack with Two-phase Specific Triggers}

\author{
Nan Luo$^1$
\and
Yuanzhang Li$^1$\and
Yajie Wang$^2$\and
Shangbo Wu$^2$\and
Yu-an Tan$^2$\And
Quanxin Zhang\textsuperscript{1,}\thanks{Corresponding author.}
\affiliations
$^1$School of Computer Science and Technology, Beijing Institute of Technology\\
$^2$School of Cyberspace Science and Technology, Beijing Institute of Technology
\emails
\{luonan, zhangqx\}@bit.edu.cn
}

\begin{document}

\maketitle
\begin{abstract}
Backdoor attacks threaten Deep Neural Networks (DNNs). Towards stealthiness, researchers propose clean-label backdoor attacks, which require the adversaries not to alter the labels of the poisoned training datasets. Clean-label settings make the attack more stealthy due to the correct image-label pairs, but some problems still exist: first, traditional methods for poisoning training data are ineffective; second, traditional triggers are not stealthy which are still perceptible. To solve these problems, we propose a two-phase and image-specific triggers generation method to enhance clean-label backdoor attacks. Our methods are (1) powerful: our triggers can both promote the two phases (i.e., the backdoor implantation and activation phase) in backdoor attacks simultaneously; (2) stealthy: our triggers are generated from each image. They are image-specific instead of fixed triggers. Extensive experiments demonstrate that our approach can achieve a fantastic attack success rate~(98.98\%) with low poisoning rate~(5\%), high stealthiness under many evaluation metrics and is resistant to backdoor defense methods.

\end{abstract}

\section{Introduction}

Deep Neural Networks (DNNs) have shown its great power to many crucial tasks which human beings may not work well such as image classification, voice recognition, auto-driving, natural language processing, etc.~\cite{chen2015deepdriving,he2016deep}. However, training models are time and resource costly. Not everyone can burden that, so people who want to use DNNs may use the third-party or public service and resource like the pre-trained models and datasets to gain a model for their specific tasks. 


Backdoor attacks become the potential threat with significant harm in such a scenario. BadNets~\cite{gu2017badnets} first revealed this attack where adversaries can implant hidden backdoors in the victim model. The backdoor model behave normally on the clean images in the inference phase. However, when adversaries activate the hidden backdoor with a trigger, the backdoor model will behave maliciously, producing the target results as adversaries expected.

Existing methods for backdoor attacks include two categories. The first one is altering-label type attack~ \cite{gu2017badnets,liu2017trojaning,liao2018backdoor,yao2019latent,li2020-imgsp-backdoor}. The adversaries will replace the correct labels of poisoned images with the target label. The second one is clean-label type attacks\cite{turner2018clean,barni2019new,saha2020hidden,quiring2020backdooring}. The adversaries will keep the correct image-label pairs of poisoned images.

In the altering-label type backdoor attacks, the mislabeled poisoned images are straightforward for the human reviewers to remove. So we focus on the clean-label backdoor attacks. However, some problems still exist: first, implanting a backdoor through under clean-label settings is more difficult. Our experiments demonstrate that former clean-label backdoor attacks CLBA~\cite{turner2018clean} may fail under high-resolution datasets. Second, former clean-label backdoor attacks' triggers are not stealthy, which are often fixed square patterns. In this case, these triggers are still easy for the human reviewers to remove because they look very abrupt upon the images. Our experiments also show such triggers is not stealthy under some evaluation metrics.

We start from effectiveness and stealthiness to consider our trigger generation methods. For the effectiveness, we consider different tasks in backdoor implantation and activation phases. In the backdoor implantation phase, inspired by CLBA~\cite{turner2018clean}, we argue that the key to implant backdoor in pretrained victim models is that the poisoned data can be sufficiently learnt. Once the victim model learns the poisoned images, it will build the connection between triggers and target labels. For example, the clean-label type backdoor attack CLBA uses an adversarial perturbation to erase poisoned images' original features to promote the victim model to learn the trigger's features. In the backdoor implantation phase, previous triggers are only effective when the backdoor is implanted. We consider building more powerful triggers that can be effective before and after backdoor implantation. Our later experiments demonstrate this type of trigger is effective.
For the stealthiness problems, we build triggers generated from each image instead of fixed triggers. They are more stealthy than the fixed triggers.

To achieve image-specific triggers which can enhance the effectiveness and stealthiness in backdoor implantation and activation phases, we utilize a U-Net autoencoder to generate triggers from each image. We propose a loss function containing target images loss (aims to enhance backdoor implantation) and non-target images (aims to enhance backdoor activation), and perceptual loss (aims to enhance the stealthiness) for the autoencoder. After obtaining a pretrained trigger generator, we use it to build poisoned images in the backdoor implantation phase and build malicious input to manipulate the backdoor model in the backdoor activation phase.

Our contributions are as follow:

\begin{itemize}
\item We propose a novel trigger generation method that can generate image-specific triggers. We train trigger generators with U-Net autoencoder architecture and well-design losses for different images.
\item We utilize this pretrained trigger generator to perform effective and stealthy backdoor attacks under clean-label settings. We compare our approach with other backdoor attacks and study our approach's performance under different parameters.
\item Extensive experiments prove our approach is effective and stealthy. We can achieve the highest attack success rate at 98.98\% with only 5\% poisoning rate, remaining benign accuracy at 97.98\%, which only drop 0.34\% compared to the clean model. Besides, Our attacks are more stealthy than other backdoor attacks under evaluation metrics; and our attacks can be resistant to backdoor defense methods.
\end{itemize}

\section{Related Work}
{\bf BadNets.}~\cite{gu2017badnets} first discovered the backdoor attack threat in the supply chain which use deep learning methods. BadNets stamps a simple trigger (i.e., one pixel or multi-pixel matrix) on the images which are selected randomly from the training datasets. For those images which are crafted, their labels should also altered into the target label. This inconsistency between the changed label and the image is quite distinct, therefore, it's easy for the human reviewers or some simple data filtering methods to remove this malicious data. Later works analyze the reason of the success of BadNet is that there a huge difference between the altered labels and the poisoned images, This difference promotes the victim model to learn connection with the trigger pattern and target label. 

\noindent {\bf Clean-Label Backdoor Attacks.}~\cite{turner2018clean} explored the backdoor attacks under higher constraints, which we refer to as clean-label settings. This work suggested a less obvious attack, the adversaries do not alter the label of the image when building malicious retraining datasets, the adversaries will first choose one attack target. Then craft images belong to this class. They first add perturbation generated by FGSM~\cite{goodfellow2014explaining} or GAN~\cite{goodfellow2014generative} on images to build the poisoned datasets and then stamp a fixed trigger on each poisoned images. They use such datasets to train a backdoor model and then use the trigger to activate the hidden backdoor. This work opens the study of clean-label type backdoor attacks, but there are some shortcuts. First, This work is operated on CIFAR-10~\cite{krizhevsky2009learning} datasets, our experiments demonstrate that this work may not work well in high-resolution images datasets like ImageNette~\cite{imagenette}. 

\begin{figure*}[t]
    \centering
    \includegraphics[width=\linewidth]{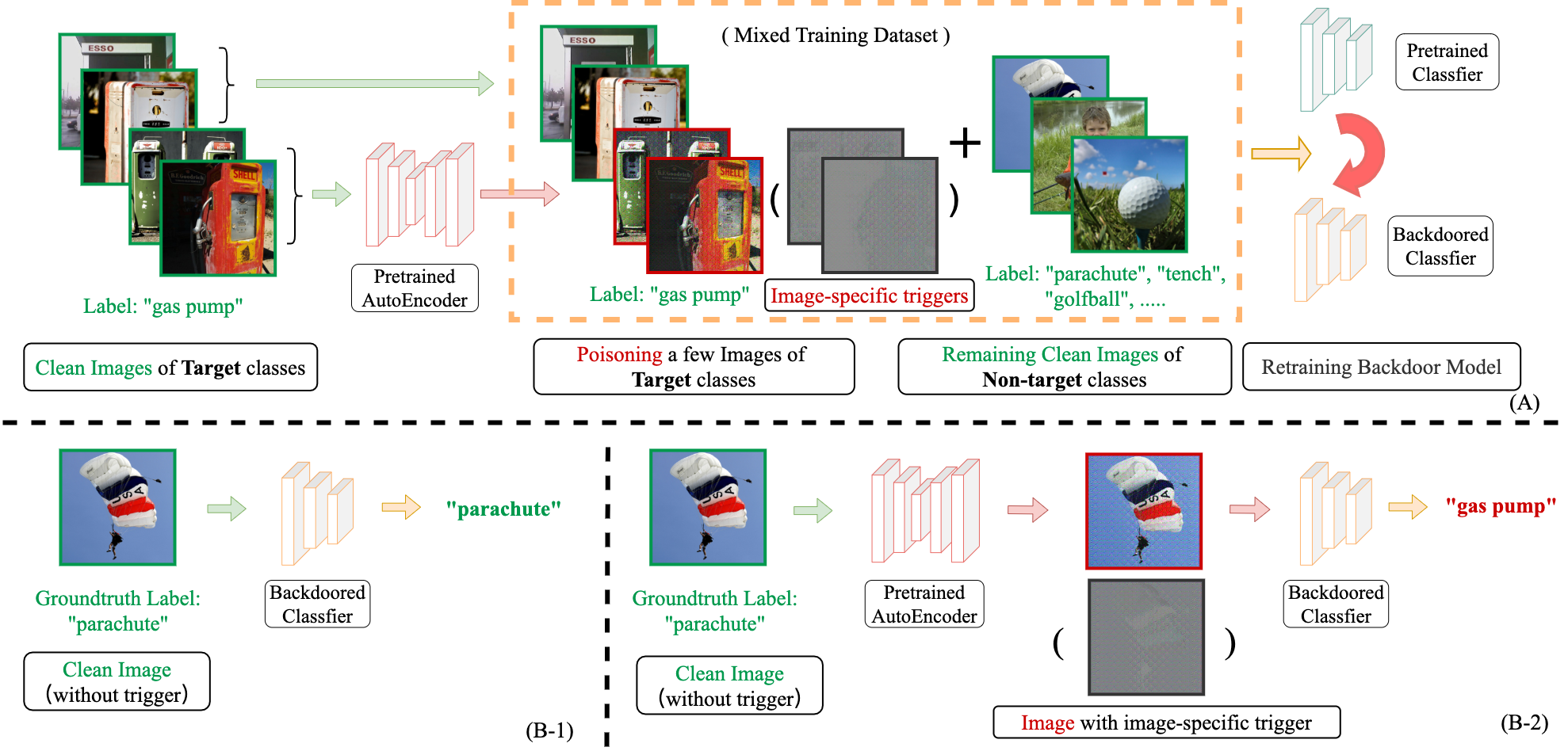}
    \caption{Overview: \textbf{(A) Backdoor Implantation: }First, we use our trigger generator \(G\)~(a U-Net autoencoder) to build a few poisoned images from the target class. Second, we mix these poisoned images with other target and non-target clean images. Third, we use the mixed datasets to train the pretrained model for a backdoor model \(f^b(\Theta,x)\). \textbf{(B) Backdoor Activation:} (B-1) shows when a clean image without trigger is input into the backdoor model, the backdoor mdoel will predict the correct result. But as (B-2) shows, if we use the trigger generator to generate a trigger from the clean image, the backdoor will be activated, and the backdoor model will predict the target result.}%
    \label{fig:attack_overview}
\end{figure*}

\noindent {\bf Generative Adversarial Perturbation}.~\cite{poursaeed2018generative} proposed generative models for creating adversarial examples, slightly perturbed images
resembling natural images but maliciously crafted to fool
pre-trained models. They presented trainable deep neural networks for transforming images to adversarial perturbations. This perturbation can achieve impressive results on high-resolution datasets such as ImageNet. Inspired by this work, in this paper, we utilize the autoencoder to build image-specific triggers from clean images which can both enhance the backdoor implantation and the backdoor activation.

\section{Methodology}


\subsection{Attack Overview}

We assume our backdoor attack happens when the adversaries have a pre-trained model \(f(\Theta,x)\), retraining dataset \(D\), where the data pair is denoted as \((x_i, y_i)\). The target class is \(y_t\) is randomly chosen as the 7th class "gas pump" in this example.

Our proposed backdoor attack consists of two phases: (A)~the backdoor implantation phase and (B)~the backdoor activation phase. Before performing backdoor attacks, we will train a trigger generator \(G\). The details of the trigger generator will be described in Section 3.4. In the backdoor implantation phase (Fig. \ref{fig:attack_overview} (A)), we first select a portion of the images in the target class and use the trigger generator \(G\) to build the poisoned data image \(x'\) with an image-specific trigger \(\Delta\). For other clean images of the target class, we keep them in the retraining datasets. Second, we mix these poisoned images with the other clean images of the rest classes. Finally, We use this mixed datasets \(D'\) to fine-tune the pretrained model \(f(\Theta,x)\) to obtain the backdoor model \(f^b(\Theta,x)\). In the backdoor activation phase, we first feed the non-targeted clean image \(x\) into the trigger generator to obtain the malicious input \(x'\) with an image-specific trigger \(\Delta\). Then we feed this malicious input \(x'\) into the backdoor model \(f^b(\Theta,x)\), the model will produce the target label. This process is shown in Fig.~\ref{fig:attack_overview} (B-2). Besides, the backdoor model \(f^b(\Theta,x)\) should produce correct prediction results for clean images without triggers. As shown in Fig.~\ref{fig:attack_overview} (B-1), the backdoor model \(f^b(\Theta,x)\) will output a correct prediction ``parachute'' when the input is clean.

\subsection{Two-phase Triggers}
To promote our backdoor attack, we want the image-specific triggers of target, and non-target images can promote both the backdoor implantation and activation phases. Specifically, in the backdoor implantation phase, we want the poisoned target data \(x'_t\) with added trigger \(\Delta\) to facilitate the classifier model \(f(\Theta,x)\) to learn the features between the trigger  \(\Delta\)  and the target class label \(y_t\). Inspired by clean-label backdoor attack~\cite{turner2018clean}, we generate specific triggers \(\Delta_t\) for the target class images \(x_t\), which can erase the features of the original image and thus enhance the backdoor implantation. i.e., we want the victim model \(f(\Theta,x)\) can not predict the correct result when the input \(x_t\) is added with a specific trigger \(\Delta_t\). We can represent this process as Eq.~(\ref{equ:tr-design-1}).
\begin{equation}
    f(\Theta, x'_t) \neq y^{t}, x'_t = x_t + \Delta_t
    \label{equ:tr-design-1} 
\end{equation}
Standard triggers are only effective when the backdoor is implanted. In our approach, we generate the triggers which can achieve some attack success rate even before backdoor implantation. Moreover, this type of trigger can be more effective after backdoor implantation. Specifically, in the backdoor activation phase, we want the crafted non-target image \(x'_{nt}\) with the added trigger \(\Delta_{nt}\) to activate the backdoor efficiently, causing the backdoor model \(f^b(\Theta,x)\) to output the target label \(y_t\) expected by the adversaries. We can represent this process as Eq.~(\ref{equ:tr-design-2}).
\begin{equation}
    f^b(\Theta, x'_{nt}) = y^{t}, x'_{nt} = x_{nt} + \Delta_{nt}
    \label{equ:tr-design-2} 
\end{equation}

\subsection{Image-specific Trigger Generator}

Inspired by GAP~\cite{poursaeed2018generative}, we use a U-Net architecture~\cite{ronneberger2015u} autoencoder as the trigger generator \(G\) to generate the image-specific triggers used in backdoor implantation phase and backdoor activation phase. This autoencoder is with skip connections between the encoder and the decoder. 

In order to obtain a trigger generator that can generate specific triggers used in different phases depend on each image, we train this trigger generator before performing the backdoor attack. This training process is described in Algorithm~\ref{alg:algorithm}. We calculate different losses depend on different types of images. i.e., we calculate the target images loss \(\mathcal{L}_t\), non-target images loss \(\mathcal{L}_{nt}\) and LPIPS loss  \(\mathcal{L}_{lpips}\), then add up these loss terms with hyperparameters (\(\alpha, \beta, \gamma\)) linearly to a total loss \(\mathcal{L}_{total}\)~(represented as Eq.~(\ref{equ:total_loss})). Then use an optimizer to minimize the loss and update the weights of the generator \(G\). 
\begin{equation}
\mathcal{L}_{total} =\alpha \mathcal{L}_t + \beta \mathcal{L}_{nt} + \gamma \mathcal{L}_{lpips}
\label{equ:total_loss}
\end{equation}
The target images loss \(\mathcal{L}_t\) is represented as Eq.~(\ref{equ:target_loss}), \(\mathcal{H}\) is the cross-entropy function, \(x'_{(t,i)}\) is the \(i\)~th crafted target image which is added by clean target image \(x_t\) and an image-specific trigger \(\Delta_{t}\), \(m\) is the total amount of target images. The trigger \(\Delta_{t}\) is generated by trigger generator \(G\) and clipped with a \(l_{\infty}\) norm limit \(\varepsilon \). \(y_{llc}\) is the least likely class of clean classifier model \(f(\Theta,x_{t})\). (as proposed by~\cite{poursaeed2018generative}). We find \(\mathcal{L}_t\) shows good results on Eq.~(\ref{equ:tr-design-1}).

\begin{equation}
\mathcal{L}_t = \sum_{i}^{m}\mathcal{H}(f(\Theta,x'_{(t,i)}),y_{llc})\ \ s.t.\ \ \| \Delta_{t} \|_{\infty} \leq \varepsilon 
\label{equ:target_loss}
\end{equation}

Similar to \(\mathcal{L}_t\), the non-target images loss \(\mathcal{L}_{nt}\) is represented as Eq.~(\ref{equ:non-target_loss}), but this loss is calculated on non-target images \(x_{(nt,i)}\), \(n\) is the total amount of target images. and we use target label \(y_t\) instead of \(y_{llc}\). We find \(\mathcal{L}_t\) shows good results 
on Eq.~(\ref{equ:tr-design-2}).

\begin{equation}
\mathcal{L}_{nt} =\sum_{i}^{n}\mathcal{H}(f(\Theta,x'_{(nt,i)}),y_{t}),\ \ s.t.\ \ \| \Delta_{nt} \|_{\infty} \leq \varepsilon 
\label{equ:non-target_loss}
\end{equation}

The LPIPS loss is represented as Eq.~(\ref{equ:lpips_loss}). LPIPS is perceptual metric proposed by~\cite{zhang2018unreasonable}, we use this loss to minimize perceptual distortion on crafted images \(x'_i\) and clean images \(x_i\) both on target images and non-target images. \(r\) is the total amount of all images, note that \(r = m+n\).

\begin{equation}
\mathcal{L}_{lpips} =\sum_{i}^{r}LPIPS(x_i,x'_i),\ \ s.t.\ \ \| \Delta_{nt} \|_{\infty} \leq \varepsilon 
\label{equ:lpips_loss}
\end{equation}

Note that, in order to keep the target image loss \(\mathcal{L}_{t}\) and the non-target image loss \(\mathcal{L}_{nt}\) converge at the same speed, the amount of non-targeted images and target images in the dataset \(D_g\) for training trigger generator is 1:1, and the amount of each class images in non-targeted images is the same. (i.e., if there are 900 images in the target class, there are 100 images in each of the other nine classes)


\begin{figure*}[t]
    \centering
    \includegraphics[width=\linewidth]{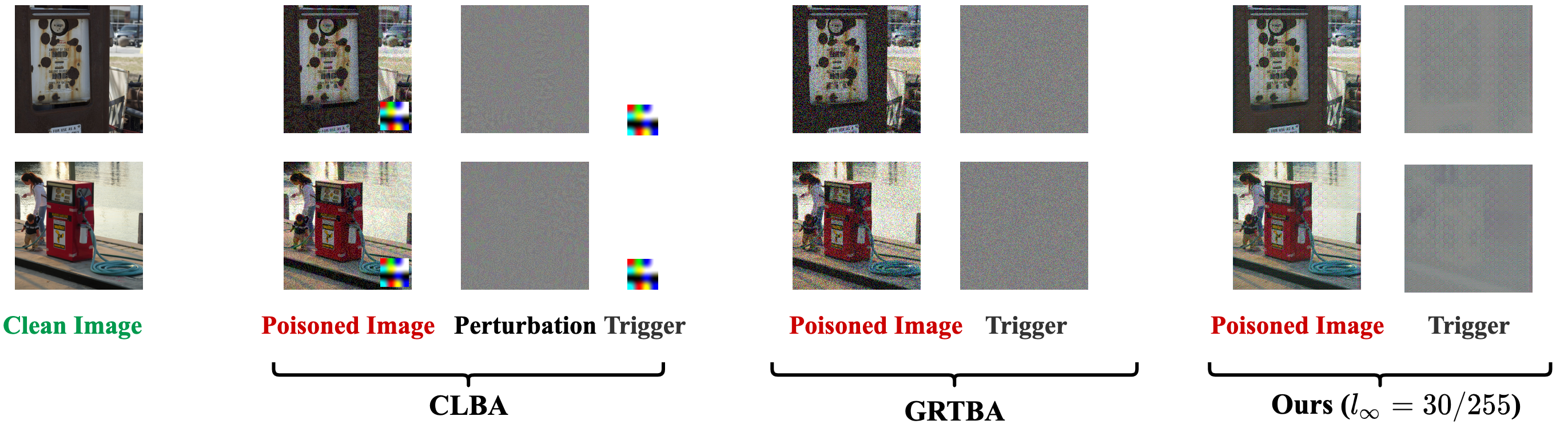}
    \caption{Visualization of poisoned images by different kinds of backdoor attacks: In Clean-Label Backdoor Attack (\textbf{CLBA}), adversaries first generate a adversarial perturbation (e.g., here using FGSM method with \(l_{\infty}=16/255\)), then add the fixed square trigger on the right-corner of images; In Global-Random-Trigger Backdoor Attack (\textbf{GRTBA}), adversaries randomly generate a global trigger~(e.g., \(l_{\infty}=40/255\) here), then add it on the whole image. \(l_{\infty}=40/255\); In \textbf{Ours} method, we use the trigger generator to generate the image-specific trigger, and then add on the whole image. Note that these image-specific triggers are generated under \(l_{\infty}=30/255\).}%
    \label{fig:exp_big}
\end{figure*}

\subsection{Backdoor Implantation and Activation}
In backdoor implantation phase: First, we use the pretrained trigger generator \(G\) to generate triggers and build poisoned images \(x'_t\) from target class images \(x_t\). We use poisoning rate \(\lambda\) to indicate ratio of the poisoned images amount to total images amount (Note that, our attacks can be effective even with poison rate \(\lambda=1\%\)). Then mix them with other clean images (including clean target class images \(x_t\) and clean non-target class images \(x_{nt}\)) to obtain a mixed retraining dataset \(D'\). Third, \(D'\) will be used by the victim to fine-tune his classifier model from a pretrained classifier model \(f(\Theta,x)\). After training, we will obtain a backdoor model \(f^b(\Theta,x)\). In the backdoor activation phase:  First, for the a non-target image \(x_{nt}\), we use the trigger generator \(G\) to generate its image-specific triggers \(\Delta_{nt}\), then add \(\Delta_{nt}\) on the \(x_{nt}\) to obtain a malicious input \(x'_{nt}\). Then we feed \(x'_{nt}\) into the backdoor model \(f^b(\Theta,x)\), the prediction will be the target class \(y_t\). 

\begin{algorithm}[tb]
\caption{Training an Image-specific Trigger Generator}
\label{alg:algorithm}
\textbf{Input}: Victim model $f(\Theta, x)$, generator $G(\Theta_g, x)$, target class \(y_t\), epoch number \(Epochs\), clean datasets \(D_g\), parameters for AdamOptimizer \(betas\);\\
\textbf{Output}: Trigger Generator;
\begin{algorithmic}[1] 
\FOR{\(epo\) in range(\(Epochs\))}
\STATE $B \leftarrow$ GetImagesBatch($D_g$) 
\STATE Set \(\mathcal{L}_t, \mathcal{L}_{nt},\mathcal{L}_{lpips} =0\)
\FOR{\((x_i, y_i)\) in \(B\)}
\STATE{\(x'_i = x_i + G(\Theta_g, x_i)\);}
\IF {\(y_i == y_t\)}
\STATE \(y_{llc} = \mathop{\arg\min} f(\Theta, x)\)
\STATE \(\mathcal{L}_t \mathrel{{+}{=}} \mathcal{H}(f(\Theta,x'_i),y_{llc})\)
\ELSE
\STATE \(\mathcal{L}_{nt} \mathrel{{+}{=}} \mathcal{H}(f(\Theta,x'_{i}),y_{t})\) 
\ENDIF
\STATE \(\mathcal{L}_{lpips} \mathrel{{+}{=}} LPIPS(x,x')\)
\ENDFOR
\STATE \(\mathcal{L}_{total} =\alpha \mathcal{L}_t + \beta \mathcal{L}_{nt} + \gamma \mathcal{L}_{lpips}\)
\STATE Update \(\Theta_g \leftarrow\) AdamOptimizer(\(\Theta_g, \mathcal{L}_{total}, betas\))
\ENDFOR
\STATE \textbf{return} \(G(\Theta_g, x)\)
\end{algorithmic}
\end{algorithm}

\section{Experiments}
\subsection{Experimental Settings}
\noindent \textbf{Dataset}:
We perform experiments on the ImageNet~\cite{deng2009imagenet} dataset, and for the convenience, we use a subset of ImageNet—Imagenette~\cite{imagenette}, which contains 10 classes, each with about 1000 training images and about 400 validation images as our dataset \(D\).

\noindent \textbf{Models}:
We choose the ResNet18~\cite{he2016deep} as the architecture of our classifier \(f(\Theta,x)\). We obtain the pretrained model from Pytorch library~\cite{paszke2019pytorch} and then modify the output of the last fc layer to 10 for our tasks.

\noindent \textbf{Training Trigger Generator}:
The datasets \(D_g\) used for training trigger generator \(G\) is split from the training set of the clean dataset \(D\). The amount of the target class (7th) images is 931, the amount of each non-target class images is 103 (total amount of non-target images is 927); The batch size, iterations and max epoch for training is set to 30, 50 and 15; for \(L_{total}\) fuction, we set hyperparameters (\(\alpha = 1, \beta = 1, \gamma =10 \)); we utilize Adam optimizer~\cite{kingma2014adam} with learning rate = 0.0002, betas = (0.5, 0.999).

\noindent \textbf{Backdoor Implantation and Activation}:
In backdoor implantation phase, for comparison with other backdoor attacks, we set poisoning rate \(\lambda\) to 5\%; for performance evaluation, we set \(\lambda\) from 1\% to 10\%. The batch size, and retraining epoch in backdoor model retraing is set to 100 and 1. We utilize Adam optimizer~\cite{kingma2014adam} with learning rate = 0.0001, betas = (0.5, 0.999)


% ...


\noindent \textbf{Evaluation Metric}: We evaluate with Fooling Rate~(FR), Attack Success Rate~(ASR) for effectiveness, and Benign Accuracy~(BA), LPIPS, PSNR and \(l_{\infty}\) for stealthiness.









\begin{table*}[t]
\begin{center}
\begin{tabular}{c|ccc|ccc}
\hline
Backdoor Phases& \multicolumn{3}{c|}{Backdoor Implantation Phase} & \multicolumn{3}{c}{Backdoor Activation Phase} \\ \hline
\diagbox[height=1.3em]{Experiments}{Metrics} & LPIPS & PSNR & \(l_{\infty}\) & ASR & BA & Drop of BA \\ \hline
Clean Model & - & - & - & - & 0.9832 & - \\
CLBA & 0.4527 & 18.34 & 242.2 & 0.1503 & \underline{0.9798} & \underline{0.0034} \\
GRTBA & 0.5409 & 16.56 & 40.00 & 0.4129 & 0.9788 & 0.0044 \\
Ours-1 & \textbf{0.2838} & \textbf{20.95} & \textbf{25.00} & \underline{0.9549} & \textbf{0.9811} & \textbf{0.0020} \\
Ours-2 & \underline{0.3295} & \underline{19.53} & \underline{30.00} & \textbf{0.9898} & \underline{0.9798} & \underline{0.0034} \\ \hline
\end{tabular} 
\caption{Evaluating different backdoor attacks (CLBA, GRTBA, Ours-1, Ours-2) on different metrics. we use LPIPS, PSNR and \(l_{\infty}\) to evaluate the stealthiness in the backdoor implantation phase, and use ASR, BA to evaluate the attack effectiveness and the infection on normal behaviours. The best result is denoted in boldface, second-best result is denoted with underline. Ours attacks can achieve higher effectiveness and stealthiness than the other two methods. 
}
\label{tab:1}
\end{center}
\end{table*}

\subsection{Compare with other Backdoor Attacks}


In this section, we demonstrate the effectiveness and stealthiness of our methods. We mainly compare our attack method with other backdoor attacks under clean-label settings, including Clean-Label Backdoor Attack (CLBA)~\cite{turner2018clean} and Global-Random-Trigger Backdoor Attack (GRTBA). CLBA is the first work to consider backdoor attacks under clean-label settings. This work first adds an adversarial perturbation to the clean image and then add a fixed square trigger on the right corner of the images. In this experiment, We utilize the FGSM as the perturbation and set \(l_{\infty}=16/255\), which can achieve FR at 95.69\%. For the trigger, we set it to 50\(\times\)50 size. GRTBA is a backdoor attack that uses a global random noise as the trigger. In this attack, adversaries first generate a random noise~(by Numpy Library~\cite{harris2020array}) of the same size as the clean images, then clip it with a \(l_{\infty}\) limit. Note that we do this experiment to demonstrate that global random trigger is not able to achieve attack success even with a higher \(l_{\infty}\) limits. In this experiment, We set the \(l_{\infty}=40/255\). For our methods,  we train two trigger generators with trigger clip limit \(l_{\infty}=25/255\) and \(l_{\infty}=30/255\). We use these two trigger generators to conduct our attack. Final, all poisoned images, relevant triggers and perturbations of CLBA, GRTBA and Ours, are visualized in Fig.(\ref{fig:exp_big}).

Table. \ref{tab:1}.  shows the results. All experiments are conducted on ImageNette, and the poisoning rate, retraining epoch is set to 5\% and 1. In terms of effectiveness, our methods can achieve high ASR (0.9898 and 0.9549 for Ours-1 and Ours-2) while CLBA and GRTBA can only achieve ASR at 0.1503 and 0.4129. Meanwhile, the BA of Ours remains very close to the clean model, the drop of BA is less than 0.004. In terms of stealthiness, Ours-1 achieve the best results on LPIPS, PSNR, and \(l_{\infty}\) and BA with a second-best ASR. Ours-2 achieve the best results on ASR
with second-best LPIPS, PSNR, and \(l_{\infty}\).

We also compare some other backdoor attacks on stealthiness, including BadNets, Clean-Label Backdoor Attacks on Video Recognition Models (CLBA-V)~\cite{zhao2020clean}, Hidden Trigger Backdoor Attack (HDTBA)~\cite{saha2020hidden}, Backdoor Attack with Sample-Specific Triggers (SSTBA)~\cite{li2020-imgsp-backdoor}. Clean label settings and image-specific trigger are two properties making backdoor attacks more stealthy. Clean-label settings play an essential role in stealthy backdoor attacks. Compare with altering-label type backdoor attacks, we think, even the triggers used in altering-label type backdoor attacks are invisible to humans, the huge inconsistency between the original features of the poisoned images and the target label is still apparent to human reviewers, then the human may remove these data, resulting in the failure of the backdoor attacks. Clean-label settings are the first-level stealthy properties. After applying clean-label settings, we can consider applying more stealthy triggers. The fixed triggers used in former clean-label type backdoor attacks are still apparent to human reviewers. Thus, we think image-specific triggers are more stealthy than the fixed triggers.
As Table.\ref{tab:2} shows, our methods both satisfy the clean-label settings and the specific trigger, that is more stealthy than other backdoor attacks.

\begin{table}[t]
\begin{center}
\resizebox{\linewidth}{!}{
\begin{tabular}{c|c|c|c}
\hline
 & Clean Label & Fixed Trigger & Specific Trigger \\ \hline
BadNets & ${\times}$ & $\surd$ & $\times$ \\
CLBA-V & $\surd$ & $\surd$ & ${\times}$ \\
HDTBA & $\surd$ & $\surd$ & ${\times}$ \\
SSTBA & ${\times}$ & ${\times}$ & $\surd$ \\
Ours & $\surd$ & ${\times}$ & $\surd$ \\ \hline
\end{tabular}
}
\caption{Compare with other backdoor attacks in terms of stealthiness. Our backdoor attack is under clean-label settings with image-specific trigger, that is more stealthy than other backdoor attacks.}
\label{tab:2}
\end{center}
\end{table}

\subsection{Effect of Poisoning Rate}
\begin{figure}[t]
    \centering
    \includegraphics[width=0.8\linewidth]{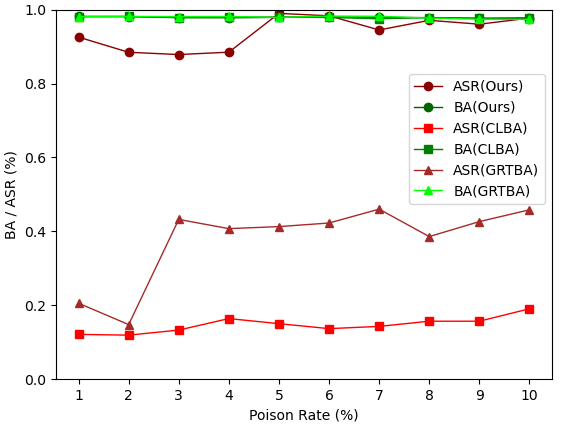}
    \caption{The effect of poisoning rate \(\lambda\) on ASR and BA of our methods, CLBA and GRTBA. Results show that our methods can achieve high ASR with low poisoning rate \(\lambda\).}%
    \label{fig:poi}
\end{figure}

The poisoning rate is an essential aspect of stealthiness. We do experiments on poisoning rate \(\lambda\) vary from 1\% to 10\% to evaluate our methods, CLBA and GRTBA. The experimental settings expect poisoning rate \(\lambda\) are the same with Section 4.2. Fig.~\ref{fig:poi} shows that our methods can achieve high ASR while maintaining high BA with a low poisoning rate~(highest ASR at 0.9898 with \(\lambda=5\%\)). CLBA and GRTBA can only achieve ASR \(\leq\) 0.1905 and 0.4601. 

\subsection{Effect of \(l_{\infty}\) limits}
We study the effect on FR and ASR with different \(l_{\infty}\) limits. And we test on the clean model and the backdoor model to demonstrate the effect of our triggers. We set \(l_{\infty}=[10/255, 15/255, 20/255, 25/255, 30/255]\).



As Table.~\ref{tab:3} shows, our triggers can achieve some FR and ASR before backdoor implantation~(highest FR at 0.7808, highest ASR at 0.4565). After backdoor implantation, we can achieve higher FR and ASR on the backdoor model than the clean model (highest FR at 0.8756, highest ASR at 0.9898). The distance of FR and ASR between the clean model and the backdoor model proves that our triggers can promote the backdoor implantation. 

Besides, as Table.~\ref{tab:3} shows, with \(l_{\infty}\) limits grows, FR and ASR also grow gradually. but there is a trade-off between stealthiness and effectiveness. the higher ASR means higher \(l_{\infty}\) for the trigger. However, we still can achieve ASR = 0.8547 with \(l_{\infty}=15/255\), the drop of ASR is only 0.1351.

\subsection{Resistance to Defense Methods}
\noindent \textbf{Resistance to Data Augmentation}:
Previous works found that data augmentation can reduce the performance of backdoor attacks. Because data augmentation may remove, destroy or alter the position of the fixed trigger.


To study this impact on CLBA, GTRBA and Ours, we performed data augmentation consisting of RandomRotation, RandomResizeCrop, RandomHorizontalFlip on the dataset for training backdoor models. We find that ASR of CLBA drops to 0.1054~(drop 0.0449); ASR of GTRBA increases to 0.6333~(increase 0.2204); ASR of Ours stays at 0.9335~(drop only 0.0214). And the BAs of these methods drop to 0.9579, 0.9600, and 0.9564, respectively.

Based on these observations, we believe that data augmentation has a destructive effect on backdoor attacks using fixed triggers. In contrast, it is less effective in defending against or even enhancing backdoor attacks using global triggers. We consider this is because the global triggers can be resistant to those data augmentation methods due to the  robustness.

\begin{table}[t]
\begin{center}
\resizebox{\linewidth}{!}{
\begin{tabular}{c|cc|cc}
\hline
Tested Model & \multicolumn{2}{c|}{Clean Model} & \multicolumn{2}{c}{Backdoor Model} \\ \hline
\diagbox[height=1.3em]{Clip Limit}{Metrics} & FR & ASR   & FR & ASR \\ \hline
\(l_{\infty}=10/255\) & 0.1778 & 0.2349 & 0.2835 & 0.3610 \\
\(l_{\infty}=15/255\) & 0.5949 & 0.4157 & 0.7587 & 0.8547 \\
\(l_{\infty}=20/255\) & 0.6848 & 0.4290 & 0.7903 & 0.8928 \\
\(l_{\infty}=25/255\) & \textbf{0.7954} & \textbf{0.4565} & 0.8705 & 0.9549 \\
\(l_{\infty}=30/255\) & 0.7808 & 0.4140 & \textbf{0.8756} & \textbf{0.9898} \\ \hline
\end{tabular}} 
\caption{Effect of different \(l_{\infty}\) limits. We test our triggers on clean models and backdoor models. Our triggers can achieve some ASR before backdoor implantation, and after backdoor implantation, our triggers can achieve higher ASR. Besides, as \(l_{\infty}\) limits grows, FR and ASR also grow gradually. The best result is denoted in boldface.}
\label{tab:3}
\end{center}
\end{table}

\begin{figure}[t]
    \centering
    \includegraphics[width=\linewidth]{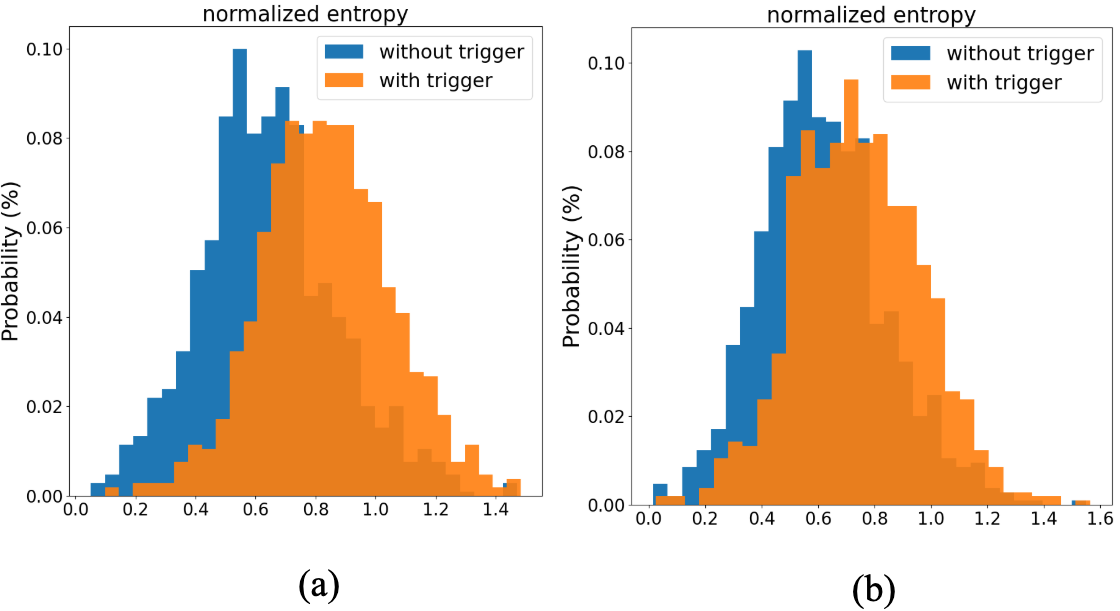}
    \caption{The entropy distribution of clean images without trigger ~(blue histogram) and images with our trigger~(red histogram). In (a), we set \(l_{\infty}=25\) to our trigger; In (b), we set \(l_{\infty}=15\) to our trigger.}%
    \label{fig:strip}
\end{figure}

\noindent \textbf{Resistance to STRIP}:
\cite{gao2019strip} propose the STRIP method to defense against backdoor attacks. STRIP superimposes various clean images on a suspicious image, then check the prediction results. If the suspicious image is with a trigger, the predictions will be invariant; if without a trig ger, the predictions will vary greatly because of the randomness.
Fig.~\ref{fig:strip} shows the entropy distribution of the clean images ~(blue histogram) and malicious images with the trigger~(red histogram). Our attacks are resistant to STRIP. The entropy distribution of images with the trigger is similar to the clean images. Fig.~\ref{fig:strip} (a) shows our attack with trigger limit \(l_{\infty}=25/255\), in this case, our attack can achieve ASR at 0.9549, the entropy median of clean images and malicious images are 0.6359 and 0.8199. Fig.~\ref{fig:strip} (b) shows our attack with a smaller trigger limit \(l_{\infty}=15/255\), in this case, our attack achieve ASR at 0.8547, the entropy median of clean images and malicious images are 0.5961 and 0.7347.

\section{Conclusion}
In this paper, we study effectiveness and stealthiness problems in clean-label type backdoor attacks. We find former clean-label backdoor attacks will fail under high-resolution datasets and their triggers are not stealthy. We propose a novel trigger generation method that can generate image-specific triggers that can promote the two phases~(i.e., the backdoor implantation and activation phases) in clean-label backdoor attacks. Extensive experiments show that our approaches can achieve a high AS with a low poisoning rate and are more stealthy than other backdoor attacks. Besides, our approaches are resistant to backdoor defense methods.


\bibliographystyle{named}
\bibliography{ijcai22}

\end{document}